\documentclass[12pt]{article}
\newcommand{\be}{\begin{equation}}
\newcommand{\ee}{\end{equation}}
\newcommand{\bea}{\begin{eqnarray}}
\newcommand{\eea}{\end{eqnarray}}
\begin{document}
\begin{titlepage}
\begin{flushright}
KEK-TH-995
\end{flushright}
\vspace{4\baselineskip}
\begin{center}{\Large\bf 
 Perturbative SO(10) Grand Unification 
}
\end{center}
\vspace{1cm}
\begin{center}
{\large 
Darwin {Chang}$^{a, b,}$
\footnote{E-Mail: chang@phys.nthu.edu.tw}, 
Takeshi {Fukuyama}$^{c,}$
\footnote{E-mail: fukuyama@se.ritsumei.ac.jp}, 
Yong-Yeon {Keum}$^{b, d,}$ 
\footnote{E-Mail: yykeum@phys.sinica.edu.tw},

Tatsuru {Kikuchi}$^{c,}$
\footnote{E-mail: rp009979@se.ritsumei.ac.jp}, 
and 
Nobuchika {Okada}$^{e,}$
\footnote{E-mail: okadan@post.kek.jp} 
}
\end{center}
\vspace{0.2cm}
\begin{center}
${}^{a}$ {\it 
Physics Department, National Tsing-Hua University, 
 Hsinchu 300, Taiwan 
}
\\[0.2cm]
${}^{b}$ {\it 
Physics Division, National Center for Theoretical Sciences, 
 Hsinchu 300, Taiwan 
}
\\[0.2cm]
${}^{c}$ {\it 
Department of Physics, Ritsumeikan University, Kusatsu, 
Shiga 525-8577, Japan
}
\\[0.2cm]
${}^{d}$ {\it 
Institute of Physics, Academia Sinica, 
Nankang 115, Taiwan
} 
\\[0.2cm]
${}^{e}$ {\it Theory Division, KEK, 
Tsukuba, Ibaraki 305-0801, Japan}
\end{center}
\vspace{1cm}
\begin{abstract}

We consider a phenomenologically viable $SO(10)$ grand 
 unification model of the unification scale $M_G$ around $10^{16} $ GeV 
 which reproduces the MSSM at low energy and allows perturbative 
calculations up to the Planck scale $M_P$ or the string scale $M_{st}$. 
Both requirements strongly restrict a choice of Higgs representations 
in a model. We propose a simple $SO(10)$ model with a set of 
Higgs representations 
$\{ 2 \times {\bf 10} + {\bf \overline{16}} + {\bf 16} + {\bf 45} \}$ 
and show its phenomenological viability. 
This model can indeed reproduce the low-energy experimental data 
relating the charged fermion masses and mixings. 
Neutrino oscillation data can be consistently 
 incorporated in the model, leading to the right-handed neutrino 
 mass scale $M_R \simeq M_G^2/M_P$. 
Furthermore, there exists a parameter region 
 which results the proton life time 
 consistent with the experimental results. 

\end{abstract}
\end{titlepage}
\newpage

\section{Introduction}

The renormalization group (RG) analysis seems 
 to favor supersymmetric (SUSY) grand unified theories (GUTs) 
 over the non-supersymmetric ones. 
In particular, with the particle contents 
 of the minimal supersymmetric standard model (MSSM), 
 the three gauge coupling constants converge at the GUT scale 
 $M_G \simeq 2 \times 10^{16}$ GeV~\cite{coupling unification I, 
 coupling unification II}. 
In addition, the recent progress in neutrino physics~\cite{PDG} 
 makes $SO(10)$ GUTs~\cite{so10} the favorite candidate 
 for grand unified theories 
 because it naturally incorporates the see-saw mechanism~\cite{see-saw} 
 that can naturally explain the lightness of the light neutrino masses. 

Recently, there has been a lot of attention paid to propose 
 and investigate a ``minimal'' $SO(10)$ model 
 with so few super multiplets that it can not only fit 
 the current sets of Standard Model data 
 and can even predict a few neutrino-related parameters 
 that experiments are yet to measured accurately. 
One example of such minimal $SO(10)$ model 
 uses the irreducible representations 
 ${\bf10}+{\bf \overline{126}}+{\bf 126}$ 
 in additional to the usual quark and lepton multiplets 
 of three ${\bf 16}_i$ $(i=1,2,3)$  and 
 only renormalizable operators~\cite{babu, matsudaetal, F-O, type II}.  

One of the main undesirable feature of this approach is that, 
 with the sizes of  the super multiplets employed, 
 they contribute such high beta function to the RG evolution 
 such that the GUT gauge coupling constant very quickly 
 blows up to infinity soon after the unification scale, $M_G$.  
For example, in the model with a set of Higgs representations 
  $\{ {\bf 10} + {\bf \overline{126}} + {\bf 126} 
 + {\bf 210} \}$~\cite{210-1, 210-2}, 
 the coupling constant diverges at $4.2 \times M_G$. 
While this cannot a priori rule out the model, 
 however it does indicate some unknown physics may take over 
 even before we reach the string scale or the Planck scale.  
One possibility to explain this run away coupling constant phenomena 
 is to argue that the string scale is actually very near the GUT scale 
 such as in some M-theories~\cite{M-Theory}. 
However, this would make the success of GUT-related phenomenology 
 more dubious, since we can not neglect non-renormalizable operators 
 originated from unknown new physics just above the GUT scale. 
It may be desirable, if achievable, 
 to keep the GUT coupling constant perturbative 
 for at least a couple of order of magnitude 
 before it reaches the (perturbative) string scale 
 $M_{st} \simeq 5 \times 10^{17}$ GeV~\cite{string scale}  
 or the (reduced) Planck scale 
 $M_P \simeq 2.4 \times 10^{18}$ GeV. 

On the other hand, the desert scenario associated 
 with the success of the MSSM  coupling constant unification 
 dictates that the GUT scale has to 
 be only about two orders of magnitude lower than the Planck scale.  
It is unavoidable that some higher dimensional operators 
 induced by the higher string scale, $M_{st}$ 
 or the Planck scale $M_P$, may play some crucial phenomenological role 
 in the analysis of the GUT models.  
This of course can make the simple GUT models much less predictive.  
However, it is unnatural also to analyze GUT models 
 pretending that the string or the Planck scales 
 are not out there not too far away.  
One reasonable strategy to pursue predictability 
 is to use only a minimal set of higher dimensional operators 
 as dictated by the requirement of 
 fitting the low energy phenomenology.  
This bottom up approach will leave it to the eventual string 
 or Planck scale physics to explain 
 why only these subsets of higher dimensional operators 
 should play important role in the GUT model analysis. 

With these perspective in mind, in this paper, 
 we propose a different approach to the $SO(10)$ unification.  
We pose the question: 
 is it possible to have realistic $SO(10)$ unification 
 with perturbative coupling constant 
 up to the Planck scale (or the string scale)?
We require the GUT model to have: \\
(1) The coupling constant unification similar to that of MSSM.  
This will require that even if there is intermediate scales 
 below the GUT scale, it will have to quite close to $M_G$.\\
(2) The GUT coupling constant remain perturbative, 
 say $\alpha_G \leq 1$, up to the Planck scale.  
While it is possible the new physics may come 
 in at the string scale lower than the Planck scale, here, 
 in the first analysis, we use Planck scale 
 because it gives stronger constraint 
 on the beta function of the GUT theory.\\ 
(3) The GUT model should fit all known low energy experimental data 
 for the Standard Model parameters 
 including CP violating phase. 
There is an issue of the role played by the yet undetermined 
 soft SUSY breaking terms. 
Here we shall assume initially that they play no role 
 in the fit to low energy parameters. 
It is partly because soft SUSY breaking sector is 
 the most uncertain part of this analysis. 
It is reasonable to leave them out 
 until it is determined later that they are needed to perfect the model. 

\section{Perturbative $SO(10)$ }

The requirement that the $SO(10)$ gauge coupling constant 
 remains perturbative up to $M_P$ imposes severe constraint 
 on the set of matter and Higgs representations we can use.  
To derive this constraint, note the solution 
 of the (one-loop) RG equation 
 for the unified gauge coupling $\alpha_G$, 
\begin{eqnarray}
\frac{1}{\alpha_G (\mu)}
=\frac{1}{\alpha_G (M_G)}-
\frac{b}{2 \pi}\log\left(\frac{\mu}{M_G}\right) \;,  
\end{eqnarray}
where $b = - b_{gauge} + b_{matter} + b_{Higgs}$ 
 is the beta function coefficient. 
Each chiral super multiplet contributes $l/2$ to $b$, 
 and each vector (gauge) multiplet contributes $3 l/2$ 
 where $l$ is the Dynkin index of the irreducible representation 
 listed in Table 1. 
{\sf
\begin{table}
\begin{center}
\begin{tabular}{|c|c|}
\hline
IRREP & $l/2$ \\
\hline \hline
{\bf 10} & 1 \\
{\bf 16} & 2 \\
{\bf 45 } & 8 \\
{\bf 54 }& 12 \\
{\bf 120 }& 28 \\
{\bf 126 }& 35 \\
{\bf 210 }& 56 \\
\hline
\end{tabular}
\caption{List of the Dynkin index for the 
$SO(10)$ irreducible representations up to the 
${\bf 210}$ dimensional one}
\end{center}
\end{table}}
For $SO(10)$ with three families, $b_{gauge}=24$ 
 and $b_{matter}=2\times 3$, 
 therefore $b = -18 + b_{Higgs}$. 
If we take the constraint and allows the coupling constant 
 to blow up at $\mu=\Lambda$, namely $1/\alpha_G (\Lambda) =0$, 
 we obtain 
\begin{eqnarray} 
  b_{Higgs} \leq 18 + {2\pi \over \ln({\Lambda \over M_G})} 
  \times {1\over \alpha(M_G)} \;.
\end{eqnarray}
In MSSM RG analysis, one typically finds $1/\alpha(M_G) \sim 24$.  
Therefore $b_{Higgs} \leq 49$, 
 if one uses $\Lambda=M_P \simeq 2.4 \times 10^{18}$ GeV, 
 the reduced Planck mass. 
If one use the stricter condition that $\alpha(M_P) =1$, 
 then the constraint becomes $b_{Higgs} \leq 48$ 
 which is about the same as before. 
Clearly to keep the couplings perturbative, 
 it is necessary to largely reduces the Higgs representations. 
It is clear from Table~1 
 that the Higgs representations 
 $\{ {\bf \overline{126}}+{\bf 126} \}$ or ${\bf 210}$ 
 are forbidden to be introduced into a model. 
Here note that ${\bf 126}$ is necessary 
 if ${\bf \overline{126}}$ is used to break symmetry 
 because of $D$-flatness condition 
 needed for preserving supersymmetry at the GUT scale.

\section{Classifying models}

Two main tasks of the Higgs representations 
 are (1) to break the $SO(10)$ gauge symmetry 
 down to the Standard Model one 
 and (2) to give fermion masses and mixings 
 consistent with all the current experimental data. 
While there are a priori many choices for Higgs representations, 
 we may pick up some Higgs representations 
 to make our model as simple as possible. 

For gauge symmetry breaking, the possible choices are:\\
(a) $\{ {\bf 45} +{\bf 54} \}$ which contribute $b_{Higgs} =20$;\
(b) $\{ {\bf \overline{16}} + {\bf 16} + {\bf 45} \}$ 
 which contributes $b_{Higgs} = 12$;\\
(c) $\{ {\bf \overline{16}}+ {\bf 16} + {\bf 54} \}$ 
 which contributes $b_{Higgs}= 16$. \\
The possibility (a) has been analyzed in the literature before. 
It certainly achieve the symmetry breaking down 
 to $G_{SM}=SU(3)\times SU(2)\times U(1)$.
For (b), a simple superpotential, 
\begin{eqnarray} 
W=M_{45} {\bf 45}_H {\bf 45}_H + 
 M_{16}  {\bf \overline{16}}_H {\bf 16}_H 
 + \lambda {\bf \overline{16}}_H {\bf 16}_H {\bf 45}_H 
\end{eqnarray} 
achieves the first task, and one can show that in this case, 
 ${\bf 45}_H$ can develop VEV that breaks 
 $SO(10) \rightarrow G_{2231}= 
 SU(2)\times SU(2)\times SU(3)\times U(1)$ and $H_{16}$ 
 can further break it to $G_{SM}$.  
Note that, in this case, 
 since the usual quark and lepton multiplets 
 also belong to a ${\bf 16}$, 
 it is necessary to impose a global symmetry like $R$-parity 
 to distinguish between the usual matter and Higgs multiplets. 
For (c), the superpotential is only 
\begin{eqnarray}
W = M_{54} {\bf 54}_H {\bf 54}_H  
 + M_{16} {\bf \overline{16}}_H {\bf 16}_H \;.
\end{eqnarray} 
Therefore symmetry breaking is not possible.  
However, there are still the potential 
 of using the higher dimensional operators 
 to help with symmetry breaking. 
We shall not treat this more complicated possibility 
 in this manuscript. 

For fermions masses, 
 there are lots of possible choices 
 for Higgs representations 
 and the situation can be more complex. 
We will define our ``minimal model'' as the one 
 that can accomplish all the above tasks 
 and contributes $b_{Higgs}$ as small as possible. 

Considering that top Yukawa coupling is of order one, 
 it is necessary to introduce, at least, 
 one Higgs representation 
 which has a renormalizable Yukawa coupling 
 with ${\bf 16}$ matters. 
Although both of ${\bf 10}$ and ${\bf \overline{126}}$ 
 can accomplish this task, 
 ${\bf \overline {126}}$ Higgs is forbidden 
 as discussed in the previous section. 
Thus, we introduce one ${\bf 10}$ Higgs into our model. 
Moreover, in order to incorporate Majorana masses 
 of right-handed neutrinos, 
 ${\bf \overline{16}}$ Higgs is necessary and the superpotential 
\begin{eqnarray} 
W= \frac{1}{M_P}\,Y_{\overline{16}}^{ij} {\bf 16}_i {\bf 16}_j 
 {\bf \overline{16}}_H {\bf \overline{16}}_H \;, 
\label{Maj}
\end{eqnarray} 
 can provide Majorana masses of right-handed neutrinos 
 through VEVs of the MSSM singlet components 
 in ${\bf \overline{16}}_H$ and ${\bf 16}_H$. 
It leads to a natural scale of the right-handed neutrino mass 
 as the one derived from the neutrino oscillation data,  
\bea
M_R \simeq \frac{M_G^2}{M_P} \;. 
\label{MR}
\eea
Throughout this paper, we assume 
 only these MSSM singlet components of 
 ${\bf \overline{16}}_H$ and ${\bf 16}_H$ develop their VEVs. 
This assumption is essential to write down the GUT mass matrix relations 
 for the charged fermions (see the next section). 
Then, the most reasonable choice of the Higgs representations 
 would be $\{ \bf 10 + 16 + \overline{16} + 45 \}$. 

With these Higgs representations, 
 the superpotential possibly relevant to the fermion masses 
 is given by (up to dimension 5 terms) 
\begin{eqnarray} 
W &=&  Y_{10}^{ij} {\bf 16}_i {\bf 16}_j {\bf 10}_H 
  +  \frac{1}{M_P} Y_{45}^{ij} {\bf 16}_i {\bf 16}_j {\bf 10}_H {\bf 45}_H 
  +  \frac{1}{M_P} Y_{\overline{16}}^{ij} {\bf 16}_i {\bf 16}_j 
     {\bf \overline{16}}_H {\bf \overline{16}}_H \;, 
\label{Yukawa1} 
\end{eqnarray} 
where the Yukawa coupling matrices 
 $Y_{10}$, $Y_{\overline{16}}$ are symmetric, 
 while $Y_{45}$ is antisymmetric. 
Here we have omitted a term proportional to 
 $ {\bf 16}_i {\bf 16}_j {\bf 16}_H {\bf 16}_H $, 
 since this is irrelevant to the fermion mass matrix 
 under the above assumption. 
One can introduce some global symmetry ($Z_N$ symmetry, for example) 
 to forbid some superpotential terms from the beginning, 
 so that the model becomes simpler. 
Such global symmetry also plays a crucial role 
 to forbid some dimension five operators 
 (such as $ {\bf 16}_i {\bf 16}_j {\bf 16}_k {\bf 16}_l/M_P$)
 in the starting Lagrangian, 
 which causes too rapid proton decay. 

In the second term in Eq.~(\ref{Yukawa1}), 
 a product ${\bf 10}_H {\bf 45}_H$ plays the same role 
 as a ${\bf 120}$ Higgs representation. 
After VEVs of the Higgs doublets in ${\bf 10}_H$ 
 and ${\bf 45}_H$ in the $B-L$ direction are developed, 
 the first two terms in Eq.~(\ref{Yukawa1}) 
 provide Dirac mass matrices of quarks and leptons. 
Note however that this model so far is obviously unrealistic 
 since it predicts the Kobayashi-Maskawa matrix being unity. 
This is because the two terms can be factorized 
 by the same ${\bf 10}_H$ 
 and, as result, the up-type quark mass matrix 
 is proportional to the down-type quark mass matrix. 
A simple way to ameliorate this problem 
 is to introduce a new ${\bf 10}$ Higgs 
 and the superpotential such as 
\begin{eqnarray} 
W  =  Y_{10}^{ij} {\bf 16}_i {\bf 16}_j {\bf 10}_1          
   +  \frac{1}{M_P}\, Y_{45}^{ij} 
     {\bf 16}_i {\bf 16}_j {\bf 10}_2 {\bf 45}_H \;, 
 \label{minimal}
\end{eqnarray}
where ${\bf 10}_1$ and ${\bf 10}_2$ are two Higgs multiplets 
 of ${\bf 10}$ representation.  
Here one can again introduce some global symmetry 
 under which ${\bf 10}_1$, ${\bf 10}_2$ and ${\bf 45}$ 
 transform differently, 
 so that the couplings of ${\bf 10}_1$ and ${\bf10}_2$ 
 are arranged as above. 
In this case, the second term plays the same role of  
 the elementary Higgs of ${\bf 120}$ representation 
 and thus this system is effectively the same as the one 
 with ${\bf 10 + 120}$ elementary Higgs multiplets.  
Then, our ``minimal model'' is defined 
 by the choice of the set of Higgs representations 
 $\{ 2 \times {\bf 10} + {\bf \overline{16}}+{\bf 16} + {\bf 45} \}$%
\footnote{
 The ``minimal'' models similar to our model 
 has been proposed by many authors~\cite{non-ren}. 
}.

\section{Fermion mass matrices and low energy data fitting} 

In the following, we use effective ${\bf 120}$ Higgs representation 
 in the analysis. 
The Yukawa couplings relevant to the Dirac mass matrices 
 are given by 
\begin{eqnarray} 
 W= Y_{10}^{ij} {\bf 16}_i {\bf 16}_j {\bf 10}_H 
    +Y_{120}^{ij} {\bf 16}_i {\bf 16}_j {\bf 120}_H \;,  
\end{eqnarray}  
where $Y_{10}$ and $Y_{120}$ are symmetric and anti-symmetric, respectively.  
Here note that $Y_{120}=Y_{45} \langle {\bf 45}_H \rangle /M_P$ 
 in the terms of the original superpotential 
 with the VEV of ${\bf 45}_H$ in the $B-L$ direction. 
Both of the Higgs multiplets ${\bf 10}_H$ and ${\bf 120}_H$ 
 include a pair of Higgs doublets in the MSSM decomposition. 
At low energy after the GUT symmetry breaking, 
 the superpotential leads to 
\begin{eqnarray} 
 W &=& (Y_{10}^{ij} H_{10}^u + Y_{120}^{ij} H_{120}^u) u^c_i q_j 
+ (Y_{10}^{ij} H_{10}^d + Y_{120}^{ij} H_{120}^d) d^c_i q_j  \nonumber \\
&+& (Y_{10}^{ij} H_{10}^u -3  Y_{120}^{ij} H_{120}^u) N_i {\ell}_j  
+ (Y_{10}^{ij} H_{10}^d -3 Y_{120}^{ij} H_{120}^d) e^c_i {\ell}_j \;, 
\end{eqnarray}  
where $H_{10}$ and $H_{120}$ correspond to the Higgs doublets 
 in ${\bf 10}_H$ an ${\bf 120}_H$, 
 which originate ${\bf 10}_1$ and ${\bf 10}_2$ 
 in the superpotential of Eq.~(\ref{minimal}). 
The factor $-3$ in the lepton sector is the results from 
 the VEV of ${\bf 45}_H$ in the $B-L$ direction, 
 and plays a crucial role so that unwanted GUT mass relations, 
 $m_e=m_d$ and $m_\mu=m_s$, is corrected. 

In order to keep the successful gauge coupling unification, 
 suppose that one pair of Higgs doublets 
 given by a linear combination of $H_{10}^{u,d}$ and $H_{120}^{u,d}$ 
 is light while the other pair is  heavy ($\simeq M_G$). 
The light Higgs doublets are identified as 
 the MSSM Higgs doublets ($H_u$ and $H_d$) and given by 
\begin{eqnarray} 
H_u &=& \widetilde{\alpha}_u  H_{10}^u 
 + \widetilde{\beta}_u  H_{120}^u \;, 
\nonumber \\
H_d &=& \widetilde{\alpha}_d  H_{10}^d  
 + \widetilde{\beta}_d  H_{120}^d  \; , 
\label{mix}
\end{eqnarray} 
where $\widetilde{\alpha}_{u,d}$ and $\widetilde{\beta}_{u,d}$ denote 
 elements of the unitary matrix 
 which rotate the flavor basis in the original model 
 into the (SUSY) mass eigenstates.  
Omitting the heavy Higgs mass eigenstates, 
 the low energy superpotential is described 
 by only the light Higgs doublets 
 $H_u$ and $H_d$ such that 
\begin{eqnarray}
W_Y &=& 
   u^c_i \left( \alpha^u  Y_{10}^{ij} 
 + \beta^u  Y_{120}^{ij} \right)  H_u \, q_j 
 + d^c_i \left( \alpha^d  Y_{10}^{ij} 
 + \beta^d  Y_{120}^{ij} \right)  H_d \, q_j  \nonumber \\ 
  &+& 
   N_i  \left( \alpha^u  Y_{10}^{ij} 
  -3 \beta^u Y_{120}^{ij} \right)  H_u \, \ell_j 
 + e^c_i  \left( \alpha^d  Y_{10}^{ij} 
  -3 \beta^d Y_{120}^{ij}  \right) H_d \, \ell_j \;, 
\label{Yukawa3}
\end{eqnarray} 
where the formulas of the inverse unitary transformation 
 of Eq.~(\ref{mix}), 
 $H_{10}^{u,d} = \alpha^{u,d} H_{u,d} + \cdots $ and 
 $H_{120}^{u,d} = \beta^{u,d} H_{u,d} + \cdots $, have been used. 

Providing the Higgs VEVs, 
 $\langle H_u \rangle = v \sin \beta$ and  
 $\langle H_d \rangle = v \cos \beta$ 
 with $v \simeq 174$ GeV, 
 the Dirac mass matrices can be read off as 
\begin{eqnarray} 
 M_u &=& c_{10} M_{10} +c_{120} M_{120} \;,  \nonumber \\ 
 M_d &=&  M_{10} +  M_{120} \;,  \nonumber \\ 
 M_D &=& c_{10} M_{10} -3 c_{120} M_{120} \;,  \nonumber \\ 
 M_e &=&  M_{10} -3  M_{120}  \;,  
\label{massmatrix}
\end{eqnarray}
where $M_u$, $M_d$, $M_D$ and $M_e$ 
 denote up-type quark, down-type quark, 
 neutrino Dirac, charged-lepton mass matrices, respectively.  
Note that all the mass matrices are described 
 by using only two basic mass matrices, 
 a symmetric $M_{10}$ and an antisymmetric $M_{120}$, 
 and two complex coefficients $c_{10}$ and $c_{120}$, 
 which are defined as 
 $M_{10}=Y_{10}   \alpha^d v \cos \beta$, 
 $M_{120}=Y_{120} \beta^d  v \cos \beta$, 
 $c_{10} = (\alpha^u/\alpha^d) \tan \beta$  
 and  
 $c_{120} = (\beta^u/\beta^d) \tan \beta$, respectively. 
 
These mass matrix formulas lead to the GUT mass matrix relation 
 among the quark and lepton mass matrices, 
\begin{eqnarray}  
M_e = c_d \left( M_d + \kappa M_u \right) \;,  
\label{GUTrelation}
\end{eqnarray} 
where 
\begin{eqnarray} 
c_d = - \frac{3 c_{10} +c_{120}}{c_{10} - c_{120}} \;, \nonumber \\
\kappa = - \frac{4}{3 c_{10} + c_{120}} \; . 
\end{eqnarray} 
For simplicity, 
 we assume that $M_{10}$ and $M_{120}$ are real 
 and pure imaginary matrices, respectively, 
 and $c_{10}$ and $c_{120}$ are both real. 
Then, all the Dirac mass matrices becomes hermitian~\cite{matsudaetal}
 and still CP violating. 
Note that, according to this assumption, 
 the number of free parameters in our model 
 are reduced into eleven in total; 
 six real parameters in $M_{10}$, 
 three real parameters in $M_{120}$, $c_{10}$ and $c_{120}$. 
On the other hand, the number of observables 
 we should fit is thirteen; 
 six quark masses, three angles and one CP-phase in the CKM matrix 
 and three charged lepton masses. 
Thus there are two predictions for observables, 
 whose values have been already known by experiments. 
Therefore, the data fitting in our model is very non-trivial. 
In the following analysis, the strange quark mass 
 and the CP-phase in the CKM matrix 
 will be regarded as two predictions in our model 
 (see the following discussion). 

Without loss of generality, we can begin with the basis 
 where $M_u$ is real and diagonal, $M_u = D_u$. 
In this basis, the hermitian matrix $M_d$ can be described 
 as $M_d = V_{CKM} D_d  V_{CKM}^\dagger$ 
 by using the CKM matrix $V_{CKM}$ and 
 the real diagonal mass matrix $D_d$%
\footnote{
 In general, $M_d = U \, D_d \, U^\dagger$ 
 by using a general unitary matrix 
 $U=e^{i \alpha} e^{i \beta T_3} e^{i \gamma T_8} 
    V_{CKM} e^{i \beta^\prime T_3} e^{i \gamma^\prime T_8}$. 
 In this paper, we adopt the diagonal phases to $0$ or $\pi$ 
 for simplicity. }. 
Considering the basis-independent quantities, 
 $\mathrm{tr}\left(M_e \right)$, 
 $\mathrm{tr}\left(M_e^2 \right)$ 
 and $\mathrm{det}\left(M_e \right)$, and eliminating $c_d$, 
 we obtain two independent equations,  
\begin{eqnarray}
\left(
\frac{\mathrm{tr} (\widetilde{M_e} )}
{m_e + m_{\mu} + m_{\tau}} \right)^2
&=& 
\frac{\mathrm{tr} (\widetilde{M_e}^2 )}
{m_e^2 + m_{\mu}^2 + m_{\tau}^2} \;,
\label{cond1} \\
\left( \frac{\mathrm{tr} (\widetilde{M_e} )}
{m_e + m_{\mu} + m_{\tau}} \right)^3
&=&
\frac{\mathrm{det} (\widetilde{M_e} )}
{m_e \; m_\mu \; m_\tau} \;,
\label{cond2}
\end{eqnarray}
where $\widetilde{M_e} \equiv 
V_{CKM} \, D_d \, V_{CKM}^\dagger + \kappa D_u$. 
With input data of six quark masses, 
 three angles and one CP-phase in the CKM matrix 
 and three charged lepton masses, 
 we solve the above equations and determine $\kappa$. 
Using the $\kappa$ determined, $c_d$ is fixed by 
\be
c_d = \frac{m_e + m_\mu + m_\tau}{\left(m_d + m_s + m_b \right) 
+ \kappa \left(m_u + m_c + m_t \right)}
\;.
\label{cd}
\ee
The original basic mass matrices, $M_{10}$ and $M_{120}$, 
 are described by 
\begin{eqnarray}
M_{10} 
&=& 
\frac{3+ c_d}{4} \;
 V_{CKM}  D_d  V_{CKM}^\dagger
+ \frac{c_d ~\kappa}{4} \;D_u \;, 
\label{M10}  \\ 
M_{120} &=&
\frac{1- c_d}{4} \;
V_{CKM}  D_d  V_{CKM}^\dagger
- \frac{c_d ~\kappa}{4} \; D_u \;. 
\label{M126} 
\end{eqnarray} 
Once the solutions $c_d$ and $\kappa$ are obtained, 
 $M_{10}$ and $M_{120}$ are completely determined. 

Note that it is a very non-trivial problem 
 to find a solution that satisfies 
 both Eqs.~(\ref{cond1}) and (\ref{cond2}) at the same time 
 with only one free parameter $\kappa$. 
In the following analysis, we vary two input data, 
 the strange quark mass ($m_s$) and the CP-phase ($\delta $)
 in the CKM matrix,  within their experimental errors, 
 so that both Eqs.~(\ref{cond1}) and (\ref{cond2}) 
 can be satisfied with the same $\kappa$ value in good accuracy. 
We can find a consistent solution 
 only if we input special values for $m_s$ and $\delta$. 
This fact indicates that the input values for $m_s$ and $\delta$ 
 we have used in our analysis are two predictions 
 in our model as mentioned above. 

Now let us solve the GUT relation and determine $c_d$ and $\kappa$. 
We follow the same strategy in~\cite{F-O}. 
Since the GUT mass matrix relation is valid only at the GUT scale, 
 we first evolve the data at the weak scale 
 to the ones at the GUT scale with given $\tan \beta$ 
 according to the renormalization group equations (RGEs)
 and use them as input data at the GUT scale. 
We take input the absolute values of the fermion masses at $M_Z$ 
 as follows (in GeV): 
\begin{eqnarray}
m_u &=& 0.00233, \quad m_c = 0.677, \quad m_t = 176, \nonumber\\
m_d &=& 0.00469, \quad m_s = 0.0747, \quad m_b = 3.00, \nonumber\\
m_e &=& 0.000487, \quad m_{\mu} = 0.103, \quad m_{\tau} = 1.76.  
\end{eqnarray}
Here the experimental values extrapolated from low energies 
 to $M_Z$ were used~\cite{Fusaoka-Koide}, 
 and we choose the signs of the input fermion masses as 
$\left(m_u, m_c, m_t \right) = \left( -,-,+ \right)$ 
and 
$\left(m_d, m_s, m_b \right) = \left( -,-,+ \right)$.  
For the CKM mixing angles and a CP-violating phase 
 in the ``standard'' parameterization, 
 we input the values measured by experiments as follows: 
\begin{eqnarray}
s_{12} = 0.2229, \quad s_{23} = 0.0412, \quad s_{13} = 0.0036, \quad
\delta = 57.6^\circ \;. 
\end{eqnarray}
Since it is very difficult to search 
 all the possible parameter region systematically, 
 we present our results for $\tan \beta=30$. 
Note that only the case of large $\tan \beta$  
 can be consistent with our original model, 
 since $Y_{120}^{ij} = Y_{45}^{ij} \langle {\bf 45} \rangle /M_P
 \sim 0.01 Y_{45}^{ij}$ 
 and only $Y_{10}^{ij}$ can be of order one. 
After the RGE running, we obtain the fermion masses and 
 the CKM mixing angles and the CP phase at the GUT scale, 
 and use them as input parameters 
 in order to solve Eqs.~(\ref{cond1}) and (\ref{cond2}). 
By putting the above data, Eq.~(\ref{cond1}) gives a solution
\be
\kappa = -0.01107011\cdots \;.
\label{k1}
\ee
On the other hand, from Eq.~(\ref{cond2}), 
 we obtain the solution 
\be
\kappa = -0.01107006\cdots \;.
\label{k2}
\ee
Since these solutions are coincide with each other in good accuracy, 
 we can regard it as the solution we seek. 
Using Eq.~(\ref{cd}) to determine $c_d$, 
 now we find a solution 
\begin{eqnarray}
\kappa &=& -0.0111 \;, \nonumber\\
c_d &=& -7.89 \;.
\end{eqnarray}
The existence of the solution means 
 that our model can reproduce the low energy experimental data 
 for the charged fermion sector.

\section{Proton decay}

The most characteristic prediction of the SUSY GUTs 
 is the proton decay. 
Normally in SUSY GUTs the proton decay process 
 through the dimension five operators 
 involving MSSM matters, mediated by the color triplet Higgsino, 
 turns out to be the dominant decay modes, 
 since the process is suppressed by only a power of 
 the Higgsino mass scale. 
Experimental lower bound on the proton decay modes 
  $p \to K^{+} \overline{\nu}$ 
  through the dimension five operators 
  is given by SuperKamiokande (SuperK)~\cite{sk}, 
\begin{eqnarray}
\tau (p \rightarrow K^{+} \overline{\nu}) \geq 
 2.2 \times 10^{33} \,\, [{\mathrm{years}}] \;.  
\end{eqnarray}
This is one of the most stringent constraints 
 in construction of phenomenologically viable SUSY GUT models. 
In fact, the minimal SUSY $SU(5)$ model has been argued 
 to be excluded from the experimental bound 
 together with the requirement 
 of the success of the three gauge coupling unification~\cite{mp, gn}. 
However note that 
 the minimal $SU(5)$ model predictions contradicts 
 against the realistic charged fermion mass spectrum,  
 and thus, strictly speaking, the model is ruled out 
 from the beginning. 
Obviously some extensions of the flavor structures in the model 
 is necessary to accommodate the realistic fermion mass spectrum. 
On the other hand, knowledge of the flavor structure is essential 
 in order to give definite predictions about 
 the proton decay processes through the dimension five operators. 
Some models in which flavor structures are extended 
 have been found to be consistent with the experiments in the context 
 of $SU(5)$ models~\cite{bps, ew} and 
 in $SO(10)$ extensions~\cite{PDso10}. 

As discussed in the previous section, 
 the charged fermion mass matrices have been completely determined
 in our model. 
Therefore, we can investigate proton decay rate 
 with only some free parameters. 
Our discussion follows~\cite{detailed}. 

The Yukawa interactions of the MSSM matter 
 with the color triplet Higgs induces the following 
 Baryon and Lepton number violating dimension five operator 
\bea
W=C_L^{ijkl} Q^i Q^j Q^k L^l \;. 
 \label{dim5}
\eea
Here the coefficients are given by the products of 
 the Yukawa coupling matrices and 
 the (effective) color triplet Higgsino mass matrix. 
In our model, 
 the coefficients are given by the products of 
 two basic Yukawa coupling matrices, $Y_{10}$ and $Y_{120}$, 
 and 
 the effective $2 \times 2$ color triplet 
 Higgsino mass matrix, $ M_C $, such as~\cite{210-2} 
\begin{eqnarray}
C_L^{ijkl} 
= 
\left(Y_{10}^{ij},~Y_{120}^{ij} \right)
\left( M_C^{-1}  \right)
\left( 
\begin{array}{c}
  Y_{10}^{k l} \\
  Y_{120}^{k l}
\end{array}  \right) \;.
\label{CL}
\end{eqnarray}

As discussed in the previous section,  
 the Yukawa coupling matrices, $Y_{10}$ and $Y_{120}$, 
 are related to the corresponding mass matrices 
 $M_{10}$ and $M_{120}$ such as 
\bea
Y_{10} &=& \frac{c_{10}}{\alpha^u  v \sin{\beta}} M_{10} \;, 
\nonumber\\
Y_{120} &=& \frac{c_{126}}{\beta^u  v \sin{\beta}} M_{120} \;, 
\eea
with $v \simeq 174 $ GeV. 
Here $\alpha^u$ and $\beta^u$ are the Higgs doublet mixing parameters
 introduced in the previous section, 
 which are restricted in the range 
 $|\alpha^u|^2 +|\beta^u|^2 \leq 1$. 
Although these parameters are irrelevant 
 to fit the low energy experimental data 
 of the charged fermion mass matrices, 
 there is a theoretical lower bound on them 
 in order for the resultant Yukawa coupling constant 
 not to exceed the perturbative regime. 
In order to obtain the most conservative values of 
 the proton decay rate, 
 we make a choice of the Yukawa coupling matrices 
 as small as possible. 
In the following analysis, we restrict the region 
 of the parameters in the range 
 $(\alpha^u)^2 +(\beta^u)^2 = 1$ 
 (we assume $\alpha^u$ and $\beta^u$ real for simplicity). 
Here we present an example of the Yukawa coupling matrices 
 with fixed $\alpha^u = 0.202$ ($\beta^u = 0.979$), 
\bea
Y_{10}
=
\left(
\begin{array}{ccc}
\begin{array}{c}
0.00261 \\
0.00485 \\
-0.00208 \\
\end{array}
\begin{array}{c}
0.00485 \\
0.0163 \\
-0.0414 \\
\end{array}
\begin{array}{c}
-0.00208 \\
-0.0414 \\
1.00 
\end{array}
\end{array}
\right),
 \label{Y10}
\eea
\bea
Y_{120}
=
\left(
\begin{array}{ccc}
\begin{array}{c}
0  \\
0.0000379 \, i \\
0.0000379 \, i \\
\end{array}
\begin{array}{c}
-0.0000379 \, i \\
0 \\
4.61 \times 10^{-6} \, i \\
\end{array}
\begin{array}{c}
-0.0000379 \, i \\
-4.61 \times 10^{-6} \, i \\
0
\end{array}
\end{array}
\right)\;. 
 \label{Y120} 
\eea
Note that the numerical smallness of $Y_{120}$ is 
 a consistency check of our scheme.  
Since it is a result of the higher dimensional operator, 
 its smallness relative to $Y_{10}$ indicated 
 that it is reasonable to ignore the even higher dimensional operator.
However, $Y_{120}$ itself does play important role 
 in fitting low energy data including CP violation. 
 
For the effective color triplet Higgsino mass matrix, 
 we assume degenerate eigenvalues being the GUT scale, 
 $M_G = 2 \times 10^{16}$ GeV, which is necessary 
 to keep the successful gauge coupling unification. 
Then, in general, we can parameterize the $2 \times 2$ 
 mass matrix as 
\begin{eqnarray}
M_C  = M_G \times U \;, 
\end{eqnarray} 
 with the unitary matrix, 
\begin{eqnarray}
U = e^{i \varphi \sigma_3} 
 \left(
 \begin{array}{cc}
 \begin{array}{c}
\cos{\theta} \\
- \sin{\theta}
\end{array}
\begin{array}{c}
\sin{\theta} \\
\cos{\theta}
\end{array} 
\end{array}
  \right) 
e^{i \varphi^\prime \sigma_3}  \;. 
\end{eqnarray}
Here we omit an over all phase since it is irrelevant 
 to calculations of the proton decay rate. 
There are four free parameters in total 
 involved in the coefficient $C_L^{ i j k l}$, 
 namely, $\alpha^u$, $\varphi$, $\varphi^\prime$ and $\theta$. 
Once these parameters are fixed, $C_L^{i j k l}$ 
 is completely determined. 

Through the same numerical analysis as in~\cite{detailed} 
 we can find a parameter region in which the proton life time 
 can be in the range consistent with the SuperK results. 
In fact, we can find a special colored Higgsino mass matrix 
 that can cancel the proton decay rate 
 through the dominant mode $p \rightarrow K^+ \overline{\nu}_\tau$. 
For example, for the Yukawa coupling matrices 
 of Eqs.~(\ref{Y10}) and (\ref{Y120}), it is found to be 
 ($\tan \beta =30$)
\bea
M_C = M_{\rm G} \times 
\left(
\begin{array}{cc}
\begin{array}{c}
-0.0681 \, i \\
-0.998 
\end{array}
\begin{array}{c}
0.998 \\
0.0681 \,i 
\end{array}
\end{array}
\right)\;,
\eea
in other words, $\theta = 1.64$ [rad], 
 $\varphi = 1.57$ [rad], and $\varphi^\prime = 0$ [rad]. 
With these parameters fixed, 
 the proton life time through the sub-dominant decay modes 
 is estimated as follows: 
\bea
\tau (p \rightarrow K^+ \overline{\nu}_e ) 
 &=& 2.5 \times 10^{35}\: {\rm [years]} \; , \\
\tau (p \rightarrow K^+ \overline{\nu}_\mu ) 
 &=& 4.3 \times 10^{33}\: {\rm [years]} \; .
\eea
In our analysis, we have taken 
 the averaged squark mass of the 1st and 2nd generations 
 as $m_{\widetilde{q}} = 10$ TeV 
 and the Wino mass as $M_{\widetilde{W}} = 500$ GeV. 
These results exceed the current experimental lower bounds. 
Therefore, our model passes the proton decay constraint 
 and is phenomenologically viable.

\section{Neutrino physics} 

In our model, the right-handed Majorana neutrino mass matrix 
 is generated by Eq.~(\ref{Maj}) through VEV 
 of ${\bf \overline{16}}_H$ in the MSSM singlet direction. 
Here $Y_{\bf \overline{16}}$ is the complex symmetric matrix 
 which has twelve free parameters in general 
 and it is nothing to do with charged fermion mass matrices. 
Therefore, through the see-saw mechanism, 
 there is enough number of free parameters 
 to fit all the current neutrino oscillation data. 
In other words, there is no prediction 
 for neutrino oscillation physics. 
However, there is an interesting feature 
 through the see-saw mechanism. 

In the basis where $M_e$ is (positive) real and diagonal, 
 the light neutrino mass matrix is given by the see-saw mechanism 
 $ M_\nu = M_D^T M_R^{-1} M_D $. 
$M_\nu$ is diagonalized by the Maki-Nakagawa-Sakata (MNS) 
 mixing matrix such as  
 $M_\nu = U_{MNS}^T {\rm{diag}}(m_1, m_2, m_3) U_{MNS}$. 
The current neutrino oscillation data 
 provide informations (but not complete) 
 for the mixing angles $\theta_{ij}$ in the MNS matrix 
 and $m_i$. 

Recall that, as discussed in Sec.~4, 
 all the elements in the neutrino Dirac mass matrix 
 can be determined in our model. 
Therefore, information of $M_R$ 
 can be extracted through the (inverse) see-saw relation,  
\bea
 M_R = M_D \; M_\nu^{-1} \; M_D^{T} \;,
\label{seesaw}
\eea 
 if the neutrino oscillation data are used as inputs. 
Since the current experimental data are insufficient 
 to fix all the elements in $M_\nu$, 
 $M_R$ can be described as a function of 
 parameters not yet undetermined by experiments, 
 $M_R = M_R( m_\ell, \theta_{13}, \delta, \beta, \gamma)$, 
 where $m_\ell$ is the lightest mass eigenvalues of 
 the light Majorana neutrino, $\delta$, $\beta$ and $\gamma$ are 
 the Dirac CP-phase and the Majorana CP-phases, respectively. 
Making some assumptions for these free parameters, 
 one can evaluate $M_R$ concretely, 
 and leads to predictions for physics related to 
 the right-handed neutrino mass matrix, 
 such as, the leptogenesis scenario~\cite{leptogenesis}. 
This direction would be worth investigating. 
We leave it for future works. 

In addition, an order estimation leads to 
 an interesting consequence. 
In our model, the neutrino Dirac mass matrix is approximately 
 the same as the up-type quark mass matrix, 
 and hence its heaviest eigenvalue is roughly 
 the same as the top quark mass. 
As already mentioned, the natural scale 
 of the right-handed neutrino mass is 
 $M_R \simeq M_G^2/M_P \simeq 10^{14}$ GeV. 
Therefore, according to the see-saw mechanism, 
 we find the heaviest mass eigenvalue of the light neutrinos 
 being of order $0.1$ eV. 
Interestingly, this value is close to $\sqrt{\Delta m_{\oplus}^2}$, 
 where $\Delta m_{\oplus}^2 \simeq 2.1 \times 10^{-3} \; {\rm eV}^2$ 
 is the atmospheric neutrino oscillation data~\cite{PDG}. 
This result indicates that our model prefers 
 the hierarchical case to the degenerate case 
 for the light neutrino mass spectrum.

\section{Conclusion}

Discovery of the neutrino masses and mixings 
 has made $SO(10)$ GUT models the favorite candidate 
 as new physics. 
Lots of $SO(10)$ GUT models have been intensively discussed. 
There are {\it a priori} many choices for the Higgs representations 
 to be introduced into a model. 
We have imposed the requirement that 
 the GUT gauge coupling should remain to be perturbative up to 
the (reduced) Planck scale or the string scale 
at which further new physics including quantum gravity takes over. 
This requirement has been found to be strong enough 
 to forbids Higgs representations higher than ${\bf 126}$ dimension. 
We have proposed a simple $SO(10)$ model 
 with a set of Higgs representations 
$\{ 2 \times {\bf 10} + {\bf \overline{16}} + {\bf 16} + {\bf 45} \}$, 
which can satisfy the requirement and 
gives the beta function coefficient of the GUT gauge coupling 
 as small as possible. 
It has been shown that the model is phenomenologically viable, 
namely all the charged fermion masses and mixings have been reproduced. 
In this realistic Yukawa couplings, the most stringent proton decay 
 processes has been suppressed. 
Also, the model can reproduce 
 the current neutrino oscillation data 
 with the right-handed neutrino mass 
 being of order $M_R \simeq M_G^2/M_P$.

\vskip 1cm

\leftline{\bf Acknowledgments}

The work of D.C. is supported 
 by National Science Council of ROC (Taiwan). 
 D.C. likes to thank the hospitality of KEK Theory Group 
 during his visit when this work was initiated 
 and the support from the KEK-NCTS international exchange program. 
 Y.Y.K., T.K. and N.O. thank the hospitality of 
 Physics Division of NCTS, Hsinchu in Taiwan 
 during their visiting. 
The work of Y.Y.K. is supported by 
 Grant-in Aid from NSC: NSC-92-2811-M-001-088 in Taiwan. 
The work of T.F. and N.O. is supported in part 
 by the Grant-in-Aid for Scientific Research 
 from the Ministry of Education, Science and Culture of Japan  
 (\#16540269, \#15740164). 
The work of T.K. was supported by the Research Fellowship 
 of the Japan Society for the Promotion of Science (\#7336). 

\vspace{.5cm}


%
\end{document}